\documentclass{emulateapj}

\def\stacksymbols #1#2#3#4{\def\theguybelow{#2}
        \def\verticalposition{\lower#3pt}
        \def\spacingwithinsymbol{\baselineskip0pt\lineskip#4pt}
        \mathrel{\mathpalette\intermediary#1}}
\def\intermediary #1#2{\verticalposition\vbox{\spacingwithinsymbol
        \everycr={}\tabskip0pt
        \halign{$\mathsurround0pt#1\hfil##\hfil$\crcr#2\crcr
                \theguybelow\crcr}}}
\def\lta{\stacksymbols{<}{\sim}{2.5}{.2}}
\def\gta{\stacksymbols{>}{\sim}{3}{.5}}

\shorttitle{Age of Elliptical Galaxies}
\shortauthors{Temi et al.}

\begin{document}

\title{Mid-Infrared Emission from Elliptical Galaxies: Sensitivity
to Stellar Age}

\author{Pasquale Temi\altaffilmark{1,2,3}, 
Fabrizio Brighenti\altaffilmark{4,5}, William G. Mathews\altaffilmark{4} }

\altaffiltext{1}{Astrophysics Branch, NASA/Ames Research Center, MS 245-6,
Moffett Field, CA 94035.}
\altaffiltext{2}{SETI Institute, Mountain View, CA 94043.}
\altaffiltext{3}{Department of Physics and Astronomy, University of Western
Ontario, London, Ontario, N6A 3K7, Canada.}
\altaffiltext{4}{University of California Observatories/Lick Observatory,
Board of Studies in Astronomy and Astrophysics,
University of California, Santa Cruz, CA 95064.}
\altaffiltext{5}{Dipartimento di Astronomia,
Universit\`a di Bologna, via Ranzani 1, Bologna 40127, Italy.}

\begin{abstract}
Mid-infrared observations 
(3.6 - 24$\mu$m) of normal giant elliptical 
galaxies with the {\it Spitzer} space telescope 
are consistent with pure populations of very old stars 
with no evidence of younger stars.
Most of the stars in giant elliptical galaxies are old but
the mean stellar age determined from Balmer absorption in
optical spectra can appear much younger due to a small admixture
of younger stars.
The mean stellar age can also be determined
from the spectral energy distribution in the mid-infrared 
which decreases with time relative to the optical emission and 
shifts to shorter wavelengths. 
The observed flux ratios 
$F_{8\mu{\rm m}}/F_{3.6\mu{\rm m}}$
and $F_{24\mu{\rm m}}/F_{3.6\mu{\rm m}}$
for elliptical galaxies with the oldest Balmer line ages 
are lower than predicted by recent models of single stellar 
populations. 
For ellipticals with the youngest Balmer line ages in our sample,
3-5 Gyrs, the flux ratios 
$F_{24\mu{\rm m}}/F_{3.6\mu{\rm m}}$ 
are identical to those of the oldest stars. 
When theoretical mid-IR spectra 
of old (12 Gyr) and young stellar populations are combined,
errors in the $F_{24\mu{\rm m}}/F_{3.6\mu{\rm m}}$ observations 
are formally inconsistent with a mass fraction of 
young stars that exceeds $\sim1$\%. 
This is less than the fraction of young stars expected 
in discussions of recent 
surveys of elliptical galaxies at higher redshifts. 
However, this inconsistancy between Balmer line ages and  
those inferred from mid-IR observations must be regarded 
as provisional until more accurate observations and  
theoretical spectra become available.
Finally, there is no evidence to date that central disks or patches 
of dust commonly visible in optical images of elliptical galaxies 
contribute sensibly to the mid-IR spectrum. 
\end{abstract}


\section{Introduction}

The mid-infrared emission from elliptical galaxies 
is sensitive to their mean stellar age.
As a single stellar population evolves, 
progressively less mass is lost from red giant stars 
with a corresponding decrease in mid-IR emission from 
photospheric and circumstellar dust 
(Bressan et al. 1998; 
Mouhcine 2002; Mouhcine \& Lancon 2002;
Piovan et al. 2003).

While the vast majority of stars in giant elliptical galaxies
are thought to be very old, 
enhanced Balmer line absorption in many galaxies 
indicates the presence of 
a small subfraction of much younger stars 
(Gonzalez 1993; Worthey 1994;
Tantalo et al. 1998; 
Trager et al. 2000a; Kuntschner et al. 2002).
A $\sim 5$\% population of young stars is consistent 
with the red colors of these galaxies and may 
result from past merging or internal star formation.
This low level of star formation could occur from time to 
time in all ellipticals, or alternatively the young stars could 
be a relict of the earlier evolution of elliptical 
galaxies when star formation was stronger.  
Such a transition from blue star-forming galaxies to 
red ellipticals with very little star formation is required to 
understand the increasing total mass of red galaxies 
that have evolved at nearly
constant color since redshift $z \sim 1$ 
(e.g. Gebhardt et al. 2003; Koo et al. 2005).
To preserve the tightness of the color-magnitude relation 
among ellipticals, this evolution-driven 
star formation must be quasi-continuous and avoid strong 
bursts, implying a broad range of stellar ages among 
the younger stars.
Large star formation episodes due to mergers 
may nevertheless occur in a few galaxies.

Piovan et al. (2003) have predicted 
a strong evolution of the mid-infrared 
spectral energy distribution (SED) for a single stellar population. 
Mid-IR observations should in principle confirm the presence 
of the small ``frosting'' of younger stars in elliptical 
galaxies required to explain the Balmer line ages. 
Our previous attempt to do this using archival 
15$\mu$m data from the {\it Infrared Space Telescope (ISO)}
was somewhat indeterminate due to the large fraction 
in that sample of 
unrepresentative elliptical galaxies containing 
large masses of dust and cold gas. 
Nevertheless several normal 
ellipticals in that sample with Balmer line ages 
of 2--5 Gyrs showed no 
evidence of young stars at 15$\mu$m
(Temi, Mathews \& Brighenti 2005).
We report here recent 3.6 to 24$\mu$m observations 
of a sample of more 
normal, recently archived elliptical galaxies 
observed with the 
{\it Spitzer Infrared Telescope} 
and with known Balmer line ages.
We find, however, essentially no mid-IR evidence for younger 
stars and a significant 
mismatch with recent predictions of mid-IR SEDs.

\section{Observations}

Table 1 lists our sample of 9 bright, reasonably isolated elliptical 
galaxies from the the {\it Spitzer}
Infrared Nearby Galaxies Survey (SINGS) legacy program and from
time allocated to Guaranteed Time Observers (PI G. Fazio,
program ID number: 69).
A wide range of Balmer line stellar ages is represented. 
These data reach a noise level unprecedented in previous space 
observations, a few
$\mu$Jy in the IRAC channels and about 30$\mu$Jy in the 24$\mu$m
MIPS channel.

The data were taken with the Infrared Array
Camera (IRAC) and the Multiband Imager Photometer (MIPS) (Fazio
et al. 2004, Rieke et al. 2004). Full coverage
imaging was obtained for all observations with additional sky
coverage to properly evaluate the background emission. Details on
the observing strategies, field coverage and integration times for
the SINGS program are described by Kennicutt et al. (2003).
For all nine galaxies data were recorded at four IRAC channels
(3.6, 4.5, 5.8, 8$\mu$m) while the 24$\mu$m MIPS channel was available
only for five galaxies in the selected sample.

We used the  Basic Calibrated Data (BCD) products from the Spitzer
Science pipeline (version 11.4) to construct mosaic
images for all objects. Pipeline reduction and post-BCD processing
using the MOPEX software package provide all necessary steps
to process individual frames: dark subtraction,
flat-fielding, mux-bleed correction, flux calibration, correction
of focal plane geometrical distortion, and cosmic ray rejection.

We used the task ELLIPSE in the IRAF data reduction package to
derive surface brightness profiles and to measure aperture
photometry. Photometry of these extended sources was performed in each
band by fitting surface brightness isophotes and also by measuring
the fluxes in suitable circular apertures around the centers.
A number of point sources were present in the final mosaiced images
at all bands, with the vast majority evident in channel 1 and 2
(3.6 and 4.5 $\mu$m). These were identified by eye as foreground stars and
other galaxy in the field and cross-checked using surveys at other
wavelengths (Digital Sky Survey and 2MASS). They were then masked out
before performing the isophotal fitting and surface photometry.
The correction for the extended emission was applied
to the fluxes as described in the Spitzer Observer's manual.
The uncertainties on the final absolute calibration are estimated
at 10\% for the four IRAC channels and 15\% for the 24$\mu$m data.

\section{Results}

Figure 1 shows an overview of the observations from Table 1 
arbitrarily normalized to the flux at 3.6$\mu$m and compared with 
the single population SEDs 
predicted by Piovan et al. (2003) for ages 3 and 12 Gyrs 
and for solar abundance.
Most of the elliptical galaxies in our sample have very similar SEDs 
between 3.6 and 24$\mu$m, 
but the trend does not closely follow the predicted SED for 
either old or young populations.
The most aberrant galaxy, NGC 1316, is no surprise since it 
is currently undergoing a spectacular merger with a gas and 
dust rich galaxy; we discuss the infrared emission from  
this galaxy in more detail in Temi et al. (2005).
For the remaining E galaxies in Figure 1 
the SEDs between 3.6 and 8$\mu$m 
are very similar to those of the dust-free
elliptical galaxies in the {\it Spitzer} observations of 
Pahre et al. (2004). 
At 24$\mu$m NGC 4472 is slightly weaker relative 
to the 3.6$\mu$m flux than the other 
three normal ellipticals at this MIPS wavelength.

In Figure 2 we plot the flux ratios $F_{8\mu{\rm m}}/F_{3.6\mu{\rm m}}$
and $F_{24\mu{\rm m}}/F_{3.6\mu{\rm m}}$ against the mean stellar 
age found from the Balmer line index, 
which are accurate to within about 20-30\%.
The errors for the mid-IR fluxes are dominated not by
observational signal to noise, but by calibration
errors that are uncertain at the present time.
The error bars for the mid-IR ratios in Figure 2 
are very conservatively estimated as follows:
The positive (negative) error is found by increasing
(decreasing) the numerator flux
by 10\% (15\% for the 24 $\mu$m flux)
and decreasing (increasing) the denominator flux by 10\%.
The solid lines in these plots are the predicted flux ratios for 
single stellar populations at the indicated age. 
Apart from NGC 1316, the highest point in both plots,
the observed ratios are consistent with no evolution 
as indicated by the horizontal dotted lines; 
galaxies with radically different Balmer line ages show 
no variation in their mid-IR spectral shapes. 
Note however that the data points of the oldest ellipticals 
are offset below the predicted flux ratios.

About 50-70\% of normal elliptical galaxies have 
small optically visible dusty clouds or disks in their cores 
(von Dokkum \& Franx 1995; Lauer et al. 2005), 
and our {\it Spitzer} archival sample is typical in this 
regard (Table 1). 
Since IRAC and MIPS flux ratios measured with different apertures, 
$R_e/8$ and $R_e/2$, are not substantially different, 
we assume that emission from these central dust clouds 
is not influencing the flux ratios plotted in Figure 2.
Nor is there any tendency for ellipticals without visible central 
dust (NGC 4649 and 6703) to have different mid-IR fluxes. 
These results are consistent with our discussion of 
the {\it ISO} sample (Temi et al. 2005) 
where we were unable to detect 15$\mu$m emission from 
the central dust clouds in elliptical galaxies.
Finally, we note that the mid-IR surface brightness
profiles for all galaxies observed in the {\it Spitzer} archival 
sample follow de Vaucouleurs profiles to a good approximation.

The mean Balmer line stellar age for the galaxies 
plotted in Figure 2 is assumed to be that of an old population 
skewed toward younger ages by a small admixture of younger 
populations. 
Suppose for simplicity that all the old stars have an 
age of 12 Gyrs and that the young stars have a single,  
much younger age.
In this case there is a degeneracy between 
the fraction by mass $f_{young}$ of the younger population
and its age.
The same mean Balmer line age corresponds either to 
a very small fraction of very young stars or a larger 
fraction of somewhat older stars (but still much less 
than 12 Gyrs). 
Using the Piovan et al. (2003) predictions as a guide,
we plot in Figure 3 the mid-IR flux ratios expected 
for four small fractions $f_{young}$ against the 
age of that population. The initial mass function (IMF) 
of the younger population is assumed to be no different than 
that of the 12 Gyrs old population, namely the Salpeter law.
While the $F_{8\mu{\rm m}}/F_{3.6\mu{\rm m}}$ 
ratios are generally insensitive to small
fractions of intermediate age ($\gta 4$ Gyr) stars, the 
predicted $F_{24\mu{\rm m}}/F_{3.6\mu{\rm m}}$ ratio can be used
to effectively detect small fractions of very young (2-3 Gyr) stars.
Although the three normal ellipticals in our sample 
with mean ages less than 5 Gyrs -- 
NGC 584, 3923 and 6703 --
have flux ratios $F_{8\mu{\rm m}}/F_{3.6\mu{\rm m}}$
and $F_{24\mu{\rm m}}/F_{3.6\mu{\rm m}}$ almost identical to those
of the old galaxies, and therefore consistent 
with no young stars, the calibration errors still 
allow some nonzero contribution. 

Suppose we ignore the offset in Figure 2 between observations 
and predicted flux ratios at $\sim$12 Gyrs 
and consider the sensitivity of the {\it Spitzer} observations 
to detect age variations within the errors of the 
$F_{8\mu{\rm m}}/F_{3.6\mu{\rm m}}$
and $F_{24\mu{\rm m}}/F_{3.6\mu{\rm m}}$ ratios. 
In both panels of Figure 3 we show two dotted horizontal 
lines separated by the height of 
typical error bars of our mid-IR observations 
normalized to the ratios at 12 Gyrs predicted by 
Piovan et al. (2003).
For example using the flux ratio evolution 
in the upper panel of Figure 3, the typical mid-IR flux 
errors 
in Figure 2 allow $F_{8\mu{\rm m}}/F_{3.6\mu{\rm m}}$ 
ratios approximately consistent with young populations 
with $f_{young} = 1,$ 3, 5 and 10\% and 
ages of $\sim 0.5$, 1, 1.4 
and 2 Gyrs respectively, assuming 12 Gyrs for the old 
population.
The estimated young population ages are the intersections 
of the upper dotted line with the curves of constant 
$f_{young}$ in Figure 3.
Clearly, the age of just the young stellar population 
must not exceed the Balmer line age which includes 
both old and young stars.
It is reassuring that 
the mean ages of the younger stellar population in 
NGC 584, 6703 and 3923 that are consistent with the 
predicted mid-IR SEDs and observational errors 
are also less than the global 
mean age for these galaxies determined from the Balmer absorption lines.

However, applying this same procedure to the 
lower panel in Figure 3, the typical errors 
for the $F_{24\mu{\rm m}}/F_{3.6\mu{\rm m}}$ observations 
allow the presence of young stars with mass 
fractions $f_{young} = 1,$ 3, 5 and 10\% having 
ages of $\sim 1.5$, 2.5, 3.3
and 4.7 Gyrs respectively, assuming again that the 
remaining stars are 12 Gyrs old. 
Consequently, errors in the observed flux ratios 
$F_{24\mu{\rm m}}/F_{3.6\mu{\rm m}}$ for NGC 584 and 6703, 
with Balmer ages of 2.8 and 4.8 Gyrs respectively,  
allow single younger populations with ages that 
do not exceed the Balmer line ages if $f_{young} \lta 3$\% 
and $f_{young} \lta 10$\% respectively.
It should be recognized that this requirement on the ages 
-- that the age of the young stellar population 
in a binary population not exceed the mean 
combined Balmer age of 
both young and old stars -- is a very conservative 
criterion for inconsistency.

For an alternative estimate for the age of the young 
population in these galaxies, we compare the age and 
mass fraction of the younger population from optical 
H$\beta$ indices with the mid-IR values in Figure 3.
The three youngest galaxies, NGC 584, 3923 and 
6703, have H$\beta$ indices of 2.1, 1.9 and 1.9 respectively
(Thomas et al. 2005).
Assuming an old population of 12 Gyrs, 
we determined the age and $f_{young}$ for the young population 
for H$\beta$ = 1.9 and 2.1 using 
the interactive program at Guy Worthey's website 
(adopting Padua evolutionary tracks and solar abundance) 
and these are shown in Figure 3 
as filled squares (H$\beta = 2.1$)
and circles (H$\beta = 1.9$).
It is seen from the lower panel that the 
H$\beta$ indices and theoretical 
$F_{24\mu{\rm m}}/F_{3.6\mu{\rm m}}$ ratios  
for both NGC 584 and 6703 
are inconsistent
unless $f_{young} \lta 1$\%.

The formal disagreement between Balmer 
and mid-IR ages and low upper limits for $f_{young}$ 
must be regarded as somewhat provisional 
because of the obvious difficulties associated with 
theoretical estimates of the mid-IR SED apparent 
in Figures 1 and 2.
We stress again that the preceding estimates of the 
young population ages allowed by the SEDs of 
Piovan et al. (2003) plotted in Figure 3 
has ignored the offset of the 
observations at 12 Gyrs in Figure 2.

It may be possible in the near future 
to reduce the uncertainty in the mid-IR age estimates 
when the remaining calibration errors in the {\it Spitzer} 
data reduction are better understood.
It is also important to increase the statistics
by observing more normal elliptical galaxies with Balmer ages in the 
3 - 5 Gyrs range. 
Finally, in view of the obvious discrepancies between the observed 
and predicted SEDs in Figure 1, 
it would also be useful to observe young clusters 
of known ages and 
with metallicities similar to those of elliptical galaxies 
to establish empirical SEDs for young single stellar populations.

\vskip.4in
We thank L. Piovan for providing electronic form of the
SSP model outputs.
Studies of hot gas and dust in elliptical galaxies
at UC Santa Cruz are supported by
NASA grants NAG 5-8409 \& ATP02-0122-0079 and NSF grants
AST-9802994 \& AST-0098351 for which we are very grateful.
FB is supported in part
by a grant MIUR/PRIN 0180903.
Part of this work is based  on observations made with the
{\it Spitzer Space Telescope}, which is operated by
Jet Propulsion Laboratory, California Institute of
Technology, under NASA contract 1407.

\clearpage

\begin{deluxetable}{lrrrrcrrrrrrr}
\tabletypesize{\tiny}
\tablewidth{0pt}
\tablecaption{Basic Properties of the sample and Integrated
  Photometry\tablenotemark{a}}
\tablecolumns{13}
\tablehead{
\colhead{Name}                              &
\colhead{Log $L_{B}$\tablenotemark{b}}  &
\colhead{$R_e$}                         &
\colhead{H$\beta$\tablenotemark{c}}         &
\colhead{age\tablenotemark{d}}         &
\colhead{central\tablenotemark{e}}        &
\colhead{$F_{3.6\mu m}$ }   &
\colhead{$F_{4.5\mu m}$ }    &
\colhead{$F_{5.8\mu m}$ }    &
\colhead{$F_{8\mu m}$ }    &
\colhead{$F_{24\mu m}$ }    &
\colhead{$F_{8\mu m}$/$F_{3.6\mu m}$}        &
\colhead{$F_{24\mu m}$/$F_{3.6\mu m}$}          \\
\colhead{NGC}                    &
\colhead{($L_{B\odot}$)}      &
\colhead{($^{\prime \prime}$)}   &
\colhead{}               &
\colhead{(Gyr)}               &
\colhead{dust?}                    &
\colhead{(mJy)}               &
\colhead{(mJy)}               &
\colhead{(mJy)}                    &
\colhead{(mJy)}                    &
\colhead{(mJy)}                    &
\colhead{}                    &
\colhead{}
}

\startdata

0584 &  10.39 & 27.4 &2.08$\pm$0.05 & 2.8  & yes,4 & 153.3 &  82.6 &  46.5 &
33.9 &
15.5 & 0.22 $\pm$0.06 & 0.10$\pm$0.03 \\
1316 &  10.91 & 80.7 &2.07$\pm$0.03 & 3.2  &yes,1,4& 902.9 & 540.7 & 335.0 &
270.7 &
221.0 & 0.30 $\pm$0.08 & 0.25$\pm$0.06 \\
3923 &  10.78 & 53.3 &1.87$\pm$0.08 & 3.3  & ...   & 334.5 & 189.6 & 109.3 &
72.8
&\nodata& 0.22 $\pm$0.05 & \nodata       \\
4472 &  10.92 & 104  &1.62$\pm$0.06 & 9.6  &yes,1,4&1046.4 & 587.9 & 350.0 &
249.4 &
82.5 & 0.24 $\pm$0.06 & 0.08$\pm$0.02 \\
4552 &  10.47 & 30.0 &1.47$\pm$0.05 & 12.4& yes,1,4& 249.2 & 145.2 &  76.3 &
55.8 &
24.6 & 0.22 $\pm$0.06 & 0.10$\pm$0.02 \\
4649 &  10.75 & 73.6 & 1.40$\pm$0.05& 14.1 &no,1,4  & 770.1 & 437.1 & 243.6 &
164.8
&\nodata& 0.21 $\pm$0.05 & \nodata       \\
5813 &  10.74 & 48.6 & 1.42$\pm$0.07& 16.6 &yes,1,2,4&143.6 &  83.2 &  46.9 &
31.5
&\nodata& 0.22 $\pm$0.06 & \nodata       \\
5846 &  11.02 & 82.6 & 1.45$\pm$0.07& 14.2 &yes,2  &  314.4 & 181.3 & 102.3 &
69.6
&\nodata& 0.22 $\pm$0.06 & \nodata       \\
6703 &  10.33 & 23.8 & 1.88$\pm$0.06& 4.8  &no,3   &   79.3 &  45.2 &  24.0 &
16.9 &
8.5 & 0.21 $\pm$0.05 & 0.11$\pm$0.03 \\
\enddata

\tablenotetext{a}{All fluxes are measured within aperture of $R_e/2$.}
\tablenotetext{b}{Luminosities and distances are calculated with
  $H_0=70$
  km s$^{-1}$ Mpc$^{-1}$.}
\tablenotetext{c}{H$\beta$ indices from Trager et al. (2000b).}
\tablenotetext{d}{Ages from optical Balmer absorption lines.}
\tablenotetext{e}{The references for optically visible dust are
(1) von Dokkum \& Franx (1995); (2) Tran et al. (2001); 
(3) Silge \& Gebhardt (2003); (4) Lauer et al. (2005).}
\end{deluxetable}

\clearpage

\begin{figure}
\plotone{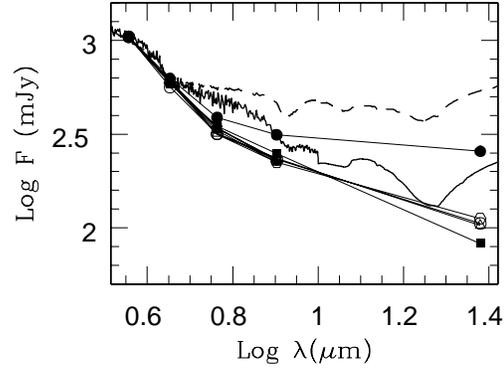}
\caption{
Observed fluxes at 3.6, 4.5, 5.8, 8 and 24$\mu$m
for
NGC 584 (open circles),
NGC 1316 (filled circles),
NGC 3923 (open squares),
NGC 4472 (filled squares),
NGC 4552 (open triangles),
NGC 4649 (filled triangles),
NGC 5813 (crosses),
NGC 5846 (stars),
and
NGC 6703 (open hexagons).
SEDs for single stellar populations
from Piovan et al. (2003) are shown at
3 Gyrs (dashed line) and
12 Gyrs (solid line).
All data and SEDs are arbitrarily normalized
at $F_{3.6\mu{\rm m}}$.
}
\end{figure}

\begin{figure}
\plotone{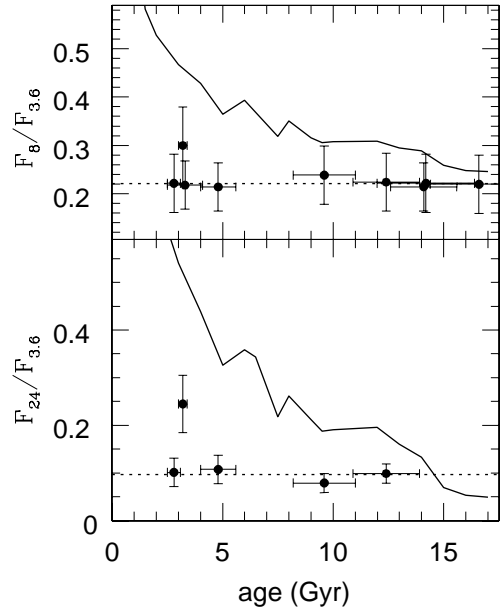}
\caption{
Observed flux ratios
$F_{8\mu{\rm m}}/F_{3.6\mu{\rm m}}$ (upper panel)
and $F_{24\mu{\rm m}}/F_{3.6\mu{\rm m}}$
(lower panel).
The solid lines show the predicted flux ratios
at each age for single stellar populations
from Piovan et al. (2003).
The dotted lines show the mean flux ratios
excluding that of NGC 1316.
}
\end{figure}

\begin{figure}
\plotone{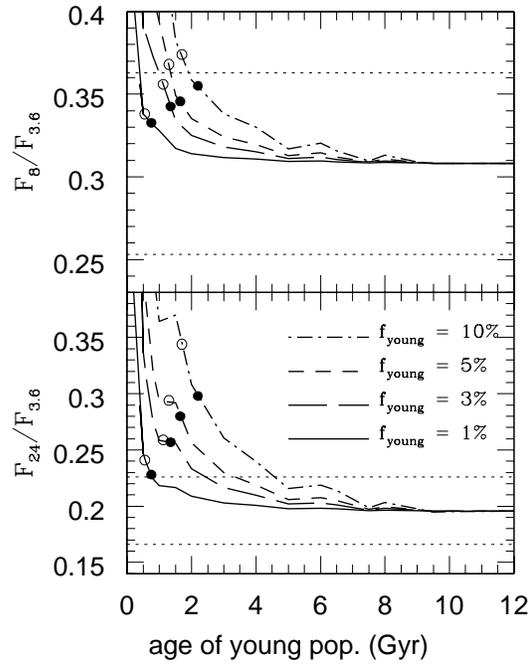}
\caption{
Flux ratios
$F_{8\mu{\rm m}}/F_{3.6\mu{\rm m}}$ (upper panel)
and $F_{24\mu{\rm m}}/F_{3.6\mu{\rm m}}$ (lower panel)
for a combination of
young stellar populations having ages on the
horizontal axis with an old stellar population
of age 12 Gyrs.
Loci of young population mass fractions $f_{young}$
of 1, 3, 5, and 10\% are shown with
solid, long dashed, short dashed, and dash-dotted
lines respectively.
The horizontal dotted lines
show the
estimated flux ratio errors of
the {\it Spitzer} observations.
The filled (empty) circles show
values of $f_{young}$ and young population age correponding to
constant H$\beta = 1.9$ (2.1).
}
\end{figure}
\end{document}